\documentstyle[preprint,aps]{revtex}
\begin{document}
\draft
 
\title{ Alouani {\it et al.} reply} 
\author{M. Alouani and J. W. Wilkins}
\address{Department of Physics, The Ohio State University, Columbus OH 43210-1168}
\author{R.C. Albers and  J. M. Wills}
\address{Los Alamos National Laboratory, Los Alamos, New Mexico 87545 }
\date{March 1994}
\maketitle

\pacs{71.20.Hk, 71.25.Tn, 71.45.Nt, 71.38.+i, 71.45.Nt}
\eject
\narrowtext

In a recent Letter \cite{alouani93}, we compared experimental results for MX (X=
Br, Cl) systems with varying M-M distances to our local density
approximation (LDA) results for Pt$_2$X$_6$(NH$_3$)$_4$ (X= Br, Cl) under 
uniaxial stress, and we showed that
{\it the key parameter} is the M-M distance.  The effect of the
different ligands on the properties of these systems is indirect; they
tune the M-M distance which is really responsible for the change in the
electronic properties.  In particular, we demonstrate that an extended
Su-schrieffer-Heeger (SSH) model with an {\it anharmonic} elastic potential
fits the entire class of Pt charge-density-wave MX materials. Thus, we have 
provided a crucial theoretical underpinning that explains the CDW MX
systems.  However, in the fit we have made a mistake in the expansion shown
by Eq. (2) of the comment. In the last term the factor $6 a^2\delta^2$ was
omitted which made it difficult for Gammel to reproduce our results. 
We have corrected this omission and found no significant effect on the 
quality of the fit to our LDA results.

The quadratic fit that  Gammel \cite{gammel} insists is the key explanation for
our LDA data does not use all the data. An  unweighted quadratic fit 
to all the points in our LDA data gives 
$\chi^2 = \sum_i (E^{LDA}_i - E^{quadratic}_i)^2$ which is
three times smaller than that of  Gammel's fit.  This unweighted fit
\cite{numbers} 
(along with a quartic fit) is shown for PtBr in our  Figs.~1 and 2 .
 Our quadratic fit strongly deviates from that of Gammel.
Gammel  used only a few LDA points at small
distortions for the  fit, which explains why Gammel curves fit
the LDA results at small dimerizations and deviates markedly at large 
dimerizations.  
From  Figs.~1 and 2 we conclude that the quadratic fit does {\it
not} explain, even at the {\it qualitative} level, the LDA and the 
experimental results \cite{okamo92,scott93} even at the {\it qualitative}
 level.  In particular, the quadratic fit does not explain: (1)  the nearly 
{\it constant} short bond-length with metal to metal (MM) distance; 
(2) the changes with respect to the MM distance in the
condensation energy (difference between the total energy at zero
dimerization and the minimum of the total energy); and (3) the
dimerization and Raman phonon frequency 
 (the predicted dimerization is half that of LDA and the
phonon frequency is 30\% smaller than that of LDA at the experimental
lattice parameter).  In short, while  the
quadratic elastic energy can fit a small region around zero dimerization, 
it is insufficient to treat the large amplitude CDW of the MX systems
which was the main physics of our original Letter \cite{comm}.

Finally we are taken to task for choosing the on-site electron-phonon 
coupling  $\beta $ to be zero.  The
complaint is that this will lead -- within the SSH two-band model -- to the
wrong ground state. In a  previous paper \cite{alouani92} we fitted an LDA
calculation to yield a value of  $\beta = 0.3 \pm 0.2 $
eV.  In the recent Letter, when we fit the model band structure 
to the LDA results under uniaxial stress we found that 
$\beta \approx 0.05$ eV, so  we set it to zero  for simplicity, 
since our aim is to discuss the physics of only CDW systems.

In summary we stress  that \underline{anharmonic}
elastic potentials are {\it essential} to explain the charge-density-wave
systematics of an entire class of MX material\cite{comm}. 

This work is supported in part by the Department of Energy (DOE) 
- Basic Energy Sciences, Division of Materials Sciences.  Supercomputer 
time was provided by the Ohio Supercomputer.

\begin{figure}
\caption{
Quadratic (solid line) and quartic (dashed line) fit \cite{numbers} to the
LDA total energy (filled circles) of PtBr as a function of dimerization 
ratio  (the ratio of the difference between the two M-X distances to 
twice the M-M
distance) for different Pt-Pt distances (from top to bottom $d_{MM}$ = 
 5.18 \AA, 5.29 \AA, 5.37 \AA,
5.45 \AA, and 5.55 \AA). 
\label{fig1} }
\end{figure}
\begin{figure}
\caption{
The calculated SSH short and long bonds with quadratic (dashed line) and 
quartic (solid line) elastic energy are compared to the 
LDA short M$^{3+\delta}$-X and long M$^{3-\delta}$-X bond
lengths for Pt$_2$Br$_6$(NH$_3$)$_4$ (open circles) as a function 
of M-M distance.  The filled triangle symbols represent experimental data for
various MX systems \cite{okamo92,scott93}.
Note that a higher polynomial expansion of the anharmonic elastic potential 
beyond the quartic will further improve the model fit to the  LDA results.
 \label{fig2}} 
\end{figure}
\widetext

\begin{references}
\bibitem{numbers} The fitted quadratic and quartic elastic energies are given by
$E^{quadratic}_{lat} = -0.04 -10.52 a + 13.47 (a^2 + \delta^2)$, and 
$E^{quartic}_{lat} = 0.05 -10.13 a + 8.10 (a^2 + \delta^2)
-5.35 (a^3 + 3a\delta^2) + 34.65 (a^4+ 6a^2\delta^2 + \delta^4)$.
\bibitem{comm} 
The need for strongly anharmonic potential to explain the MX systems is 
further demonstrated by recent calculations on PtCl compounds.
We have found that even a quartic fit is insufficient, and that only a fit 
up to the sixth order in the dimerization accurately describes the LDA 
results.
\end{references}
\end{document}